\documentclass[final,letter,authoryear,12pt,times]{elsarticle}
\usepackage{amssymb,amsmath,stmaryrd,footmisc,natbib}
\usepackage{color}
\usepackage{graphicx}
    \makeatletter
    \def\ps@pprintTitle{%
       \let\@oddhead\@empty
       \let\@evenhead\@empty
       \let\@oddfoot\@empty
       \let\@evenfoot\@oddfoot
    }
    \makeatother

\begin{document}

\begin{frontmatter}
\title{Dynamic enlargement of a hole in a sheet: crater formation and propagation of cylindrical   shock waves }

\author[me,cee]{Tal Cohen\corref{cor1}}
\ead{talco@mit.edu}

\address[me]{Department of Mechanical Engineering, Massachusetts Institute of Technology, Cambridge, MA 02139}

\address[cee]{Department of Civil and Environmental Engineering, Massachusetts Institute of Technology, Cambridge, MA 02139}

\cortext[cor1]{Corresponding author.}

\begin{abstract}

\noindent Predicting the shape of a  crater formed by high velocity impact is of interest in several fields. It can aid in design of more efficient protective structures, in forensic analysis of bullet holes, and in understanding the effects of meteorite impact in both space systems and in extreme geological events. In this paper we present, for the first time, a complete  theoretical solution of the dynamic plane-stress problem. We consider the steady-state expansion of a cylindrical hole in a strain hardening elastoplastic  sheet and find that  a self-similar field emerges if the  `specific cavitation energy' is constant. It is shown that at the quasistatic limit this solution reduces to available classical solutions, while at  high expansion velocities shock waves can appear. Investigation of the constitutive sensitivities of the expansion field is conducted and compared with available results for the spherical field which is commonly applied to predict resistance to high velocity penetration. It is shown that shock waves appear at significantly lower expansion velocities, in the plane-stress deformation pattern, for  which material compressibility is found to have a negligible effect. This insensitivity can be  taken advantage of in the future for design of light weight protective layers by incorporating porosity.

 
\end{abstract}


\end{frontmatter}
\section{Introduction}
\noindent In 1947 G.\ I.\  Taylor reports on an investigation on  the enlargement of a circular hole in a thin plastic sheet \citep{Taylor1948}. Considering a perfectly plastic material, Taylor combines the Tresca yield condition with the  Mises flow rule to arrive at what we would  refer to today as a non-associated plasticity model, and studies the quasistatic expansion process.  He compares his results with those from an  unpublished communication with Hance A. Bethe\footnote{This communication is apparently an extension of  earlier work by  \cite{bethe1941attempt}. } and concludes that   Bethe's model is \textit{``inconsistent with any theory of plasticity"}. Later, R.\  \cite{Hill1950} considers the same problem and subjects Taylor's work to similar critisism. He claims that \textit{``Taylor's method of integration was insufficiently accurate"} and that \textit{``the almost exact agreement... found experimentally by Taylor is certainly fortuitous"}. All three authors considered situations in which the hole expansion process  occurs symmetrically under plane stress conditions, such that \textit{``the metal near the hole piles up into a thickened crater"}. To examine if the symmetry assumption holds,   Taylor conducted experiments in lead sheets. The expansion was achieved by a series of tapering sections of steel cones  that penetrated the sheet while being rotated. Images of two representative samples where shown \citep{Taylor1948}, and have been included in  Appendix A of this paper.  

The controversy between Bethe, Taylor, and Hill is revisited in more detail by \cite{MCD2010}, and resolved by employing modern constitutive models and integration methods. It revolves around the values of the radial   pressure applied at the hole $(p)$, and the deformed thickness of the `crater' $(h)   $. These two values combine to predict the work invested in the quasistatic expansion process, which is proportional to their product $(ph)$. More recently, Cohen et al. have shown that, in fact, both $p$ and $h$ can become singular at the cavity wall, while their multiplication arrives at an asymptotic value  \citep{CD_PorousCavitation,CD2010,CMD2009}. Accordingly, they define the `specific cavitation energy' - $s_{c}=ph/h_0$, with $h_0$ representing the initial thickness of the sheet. This asymptotic quantity  can be interpreted as the incremental work needed to create a new unit of hole volume, during self-similar expansion. It provides a measure for the resistance of a solid to penetration.       

Motivated by the wartime need for prediction of penetration resistance of protective metal targets, the studies by Bethe, Taylor, and Hill,  provided a meaningful estimation of the energy invested in the perforation process. However, their results did not account for the dynamics of the process and, in particular, the role of inertia in resisting high velocity perforation. In fact, it is during the same years that the first theories     of dynamic propagation of plastic deformation were being formulated.  Perhaps most notable is the work of   \cite{Karman1950} that considered tensile impact of a bar.  
Quite surprisingly, to this day, the  fundamental problem  and  the quasistatic  solutions presented by Bethe, Taylor, and Hill, have not been extended to dynamic expansion. This is despite the fact that its solution can extend to explain several additional phenomena including  extreme geophysical events, such as meteorite impact \citep{melosh2005planetary,Goodier1965}, volcanic eruptions and earthquakes \citep{gudmundsson2016mechanics,gudmundsson2014mechanics}, and the growing threat of orbital debris on space operations \citep{johnson2010orbital,SMIRNOV20091796,hopkins1970effects,HOSSEINI20061006}.

In absence of  solutions for the dynamic expansion problem in plane-stress conditions, \cite{Goodier1965}, followed the notion of \cite{ BHM1945}, to   determine the indentation resistance of metal plates by employing the  spherically symmetric cavity expansion   field. To account for inertia, he used solutions by  \cite{Hill1950} and  \cite{Hopkins1960}.
This methodology has been shown to provide   good prediction for deep penetration problems in various materials \citep{forrestal1995penetration,forrestal1997spherical,masri2010estimation,chen2002deep,ben2005ballistic,warren1998effects,johnsen2018cylindrical,vorobiev2007simulation, masri2009deep, gabi2013high}, and over the years has developed into an active  research discipline.
For thin sheets, other studies continue to employ the quasistatic plane-stress expansion field pioneered by Bethe, Taylor, and Hill, as recently reviewed by \cite{ryan2018scaling}. In particular, models based on the `specific cavitation energy' \citep{cohen2010ballistic,masri2014effect,masri2015ballistically,CD_PorousCavitation,CD2010,CMD2009}, have been identified  as providing the best performance. 

In this paper we present, for the first time, a complete solution of the dynamic plane-stress problem, which reduces to the solutions by Bethe, Taylor, and Hill at the quasistatic limit. This solution is achieved by identifying that when a constant `specific cavitation energy' applied to the sample, is larger than the quasistatic value $(s_{d}>s_c)$, the cylindrical hole will expand in a self-similar manner at a constant velocity. Additionally, it is shown that if the expansion velocity exceeds a critical value, a cylindrical shock wave will form. Solutions of the entire field,  including the shock discontinuity are obtained and sensitivity to the constitutive properties  (i.e. yield stress, elastic compressibility, and strain hardening) are examined.   In the next section we begin by presenting the problem setting. Then, in section 3,  the governing equations are derived.  In section 4, the solution procedure is presented, including derivation of an analytical solution for the elastic zone. In section 5 we present results and provide a discussion in comparison with available results for the quasistatic field, and for the dynamic spherical cavity expansion field. Finally, we conclude in section 6, and discuss directions for future work.

\section{Problem setting and the self-similar field}
\begin{figure}[!h]
\centering\includegraphics[width=5.3in]{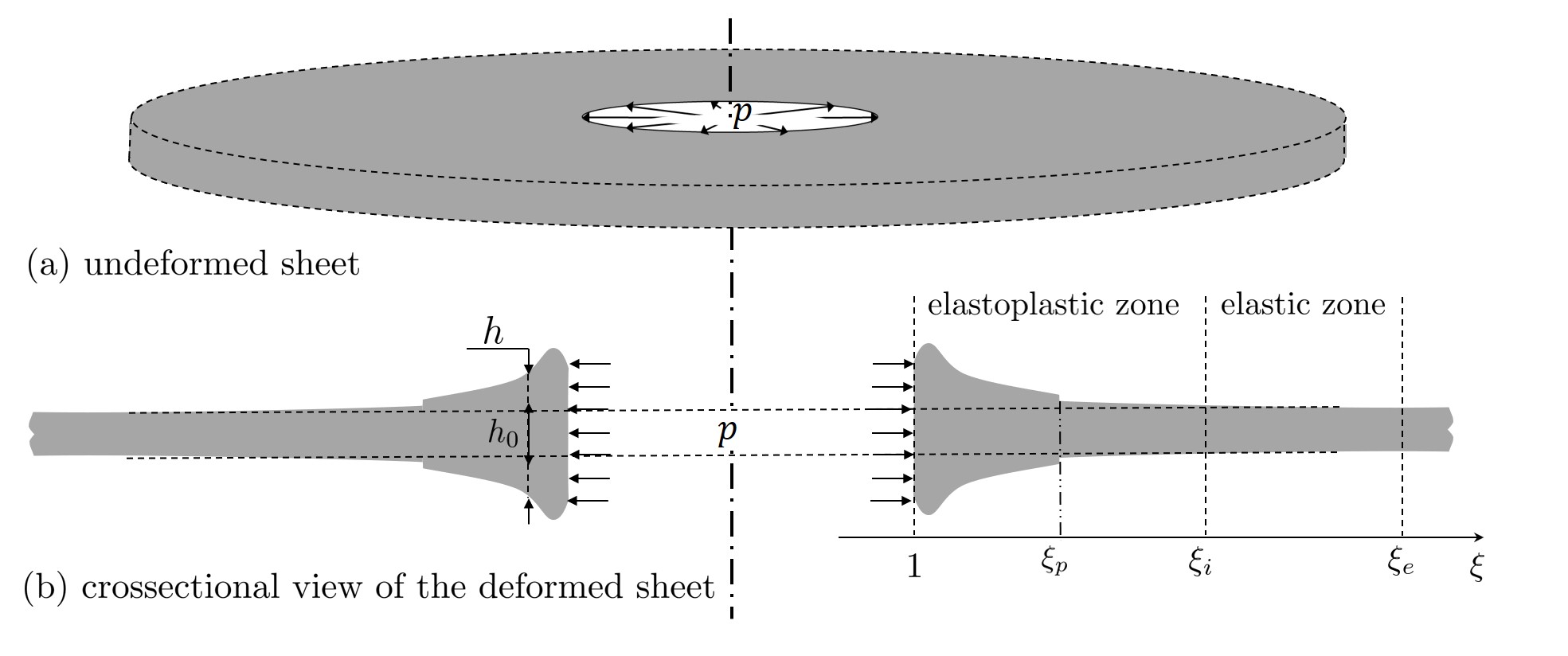}
\caption{Schematic illustration of the sheet in both its (a) undeformed configuration, and (b) deformed configuration, shown by a cross-section. The cavity wall is shown at the normalized radial location $\xi=r/a=1$, a plastic shock may appear in the elastoplastic zone $(1 \leqslant\xi\leqslant\xi_i )$ and its  normalized radial location is denoted by $\xi=\xi_p$. An elastic precursor, at $\xi=\xi_e$ propagates into the undisturbed material followed by an elastic zone $(\xi_i \leqslant\xi\leqslant\xi_e )$. The interface between the elastic and elastoplastic zones is denoted by $\xi_i$.   }
\label{illus}
\end{figure}

\noindent Consider a cylindrical cavity embedded in a plane sheet being expanded radially at  constant expansion velocity $\dot a$ by application of internal (dimensionless) pressure $p$,  as illustrated in Fig.\ \ref{illus}. 
The sheet, of undeformed thickness $h_0$,  spans indefinitely in the plane, hence disturbances are not reflected back towards the cavity.  It will be shown that in this setting, assuming plane-stress conditions and permitting changes in thickness,  a  self-similar field emerges in which the `specific cavitation energy' \citep{CD_PorousCavitation,CD2010,CMD2009}  invested in  expanding the cavity remains constant in time, although the applied pressure and deformed thickness at the cavity wall  become singular. 

Assuming  symmetry about the axis of the cylindrical cavity, the material velocity $v$ is in the radial direction and we denote the current and reference radial coordinates by $r$\ and $R$, respectively. A self-similar field must be  independent of the current radius of the cavity $a$, and thus, if it exists, all field variables can be written as functions of  a single independent variable, the dimensionless radial coordinate\begin{equation}
\xi=\frac{r}{a}~.\label{xi}
\end{equation} 

As  was previously suggested for the spherical field in \cite{DF1997} we may  define the dimensionless velocity along the  radial coordinate by \begin{equation}
V=\frac{v}{\dot a}~.\label{V}
\end{equation}Hence, at the cavity wall $(r=a)$, we have both $\xi=1$ and $V=1$.
We can now write the useful transformations
 \begin{equation}
\frac{d()}{dr}=()'~\frac{1}{a},~~~\dot{()}=()'(V-\xi)\frac{\dot a}{a}~~~\text{where}~~~()'=\frac{d()}{d\xi}~,\label{drdt}
\end{equation}for the radial and time differentiation of a field variable in the self-similar state of cavity expansion, respectively.

As shown for the spherical cavity expansion field \citep{CMD2010,CD_JAM2012,DM2004}, for materials with a definite yield point, in absence of remote loading, we expect  that at the steady-state an elastic precursor, at $\xi_e$,  propagates into the undeformed stationary regime, followed by an elastoplastic interface, at  $ \xi_i$, that separates the elastic region from the elastoplastic region, in which the material has yielded. It was shown in \citep{CMD2010,CD_JAM2012} that in the spherical setting, there exists a critical expansion velocity $\dot a$, at which a plastic shock wave may appear $(\xi_p$). In the present study we investigate the analogous field in a plane-stress deformation pattern, hence we have an additional field variable; the current thickness of the sheet, denoted by $h$.\ The propagating singularity surface, identified as a  shock wave, implies a finite jump in field variables and as such induces a localized jump in the thickness of the sheet. A similar geometrical discontinuity was obtained by  \cite{Knowles2002}, in the investigation of shock wave propagation due to tensile impact of rubber.

In the next sections we will show that  a self-similar cavity expansion field  exists in the plane-stressed deformation pattern. It is associated with a constant level of energy input invested in hole expansion (i.e. specific cavitation energy), while the applied radial pressure at the cavity wall is not finite, in contrast to  the  extensively studied spherical pattern.

\section{Governing equations}

\noindent \textbf{Continuity.} We begin by writing the continuity equations in the most general form. This will be useful later in deriving the jump conditions across a shock discontinuity. 

In the plane-stress deformation pattern, though changes in thickness of the sheet are permitted, motion is assumed to occur only in the plane. Conservation of momentum in the axially symmetric field is thus written for  radial motion\begin{equation}
\frac{\partial(h\bar\sigma_r)}{\partial r}+\frac{h}{r}(\bar\sigma_r-\bar\sigma_\theta)=  \frac{\partial( \rho hv)}{\partial t}+\frac{1}{r^{}}\frac{\partial(r\rho hv^{2})}{\partial r}~. \label{em0}
\end{equation}which employs the underlying assumption, that the field is uniform through the thickness of the sheet. Here  $\rho$ is the current density and $(\bar\sigma_r, \bar\sigma_\theta)$ are, respectively, the radial and circumferential    principal stress components. In the following formulation we will remove the bar notation to denote  the dimensionless values, henceforth all stress components in the formulation have been nondimensionalized  with respect to the elastic modulus $E.$   

The second continuity equation requires conservation of mass 
\begin{equation}
\frac{\partial( \rho h)}{\partial t}+\frac{1}{r}\frac{\partial(r\rho hv)}{\partial r}=0~.\label{cm0}
\end{equation}
By combining equations (\ref{em0}) and (\ref{cm0}) we may rewrite the equation of motion in the more compact  Navier-Stokes form   
\begin{equation}
\frac{\partial(h\sigma_r)}{\partial r}+\frac{h}{r}(\sigma_r-\sigma_\theta)=\frac{\rho}{E} h\dot v~.\label{em01}
\end{equation}We  now insert relations  (\ref{xi}) - (\ref{drdt}) into (\ref{cm0}) and (\ref{em01}) to arrive at the self-similar versions  of the equation of motion \begin{equation}
(h\sigma_r)'~+\frac{h}{\xi}(\sigma_r-\sigma_\theta)=m^2\frac{\rho}{\rho_0} h(V-\xi)V',\label{em1}
\end{equation}and conservation of mass \begin{equation}
V'+\frac{V}{\xi}+(V-\xi)\left(\frac{ h'}{ h}+\frac{\rho '}{\rho }\right)=0 \ ,\label{cm1}
\end{equation}respectively. Here we have used the dimensionless expansion velocity  \begin{equation}
m=\frac{\dot a}{C_{E}}~. \label{m}
\end{equation}where $C_E=\sqrt{E/\rho_0}$ is the wave velocity in a long elastic rod and $\rho_0$ is the density of the material in its undeformed state.
 
\vspace{0.3cm}

\vspace{0.3cm}

\noindent \textbf{Constitutive response.} Following the standard principles of Mises flow theory plasticity, we write the dimensionless Mises effective stress in the present plane-stress field as\begin{equation}
\sigma_e^2=\sigma_r^2-\sigma_r\sigma_\theta+\sigma_\theta^2~. \label{mises}
\end{equation}Employing an associated flow rule, we take the total logarithmic strain rate as the sum of a Hookean hypoelastic part and a plastic part, and invoke the principle of plastic power equivalence, which for the present deformation pattern,  leads to a set of three scalar constitutive equations
\begin{equation}
\begin{cases}
\dot\varepsilon_r&=\dot \sigma_r-\nu\dot\sigma_\theta+\eta_r\dot\varepsilon_p\ ,
\\\dot\varepsilon_\theta&=\dot \sigma_\theta-\nu\dot\sigma_r+\eta_\theta\dot\varepsilon_p\ ,\\
\dot\varepsilon_z&=-\nu(\dot \sigma_r+\dot\sigma_\theta)-(\eta_r+\eta_\theta)\dot\varepsilon_p~,\label{const0}
\end{cases}
\end{equation}  
where, for compactness,  
we have defined\begin{equation}
\eta_r=\frac{2\sigma_r-\sigma_\theta}{2\sigma_e},~~~\eta_\theta=\frac{2\sigma_\theta-\sigma_r}{2\sigma_e}~.\label{eret}
\end{equation} Here $(\varepsilon_r,\varepsilon_\theta,\varepsilon_z)$ are the radial, circumferential and out of plane logarithmic strain components, respectively,  $\nu$ is the Poisson's ratio, and $\varepsilon_p$ is the effective (logarithmic) plastic strain; a known function of the effective stress. We limit our analysis to rate independent material response and neglect thermal effects. Hence, at this point in the formulation it is most instructive to proceed with an arbitrary hardening relation $\varepsilon_p=\varepsilon_p(\sigma_e)$, which can be determined by the standard tension test for a material of interest. 

\vspace{0.2cm}
\noindent \textbf{Strain kinematics.} The logarithmic strain components in the present deformation pattern can be written as \begin{equation}
 \varepsilon_r=\ln\left(\frac{\partial r}{\partial R}\right),~~~\varepsilon_\theta=\ln\left(\frac{ r}{ R}\right),~~~ \varepsilon_z=\ln\left(\frac{h}{h_{0}}\right)~.\label{eretez}
\end{equation}
and thus the dilatation is\begin{equation}
\varepsilon_r+\varepsilon_\theta+\varepsilon_z=\ln\left(\frac{\rho_{0}}{\rho}\right)~.
\end{equation}which is an integrated form of the continuity equation \eqref{cm1}. On the other hand, by adding relations (\ref{const0}) and performing straight forward integration over time we find that \begin{equation}
\varepsilon_r+\varepsilon_\theta+\varepsilon_z=3\beta\sigma_h~,
\end{equation}where the plastic branch does not participate due to plastic incompressibility. Here $\sigma_h=(\sigma_r+\sigma_\theta)/3$ is the hydrostatic stress, and $\beta=1-2\nu$ is the compressibility parameter. Combining the above two relations yields the useful result which is independent of the loading path\begin{equation}
\frac{\rho}{\rho_0}=e^{-3\beta\sigma_h}.\label{rro}
\end{equation} 

By differentiating the strain components (\ref{eretez})  we have the strain rates \begin{equation}
\dot \varepsilon_r=\frac{dv}{dr},~~~\dot \varepsilon_\theta=\frac{v}{r},~~~\dot\varepsilon_z=\frac{\dot h}{h}\ ,
\end{equation}which,   in combination with the self-similar field relations  (\ref{xi})-(\ref{drdt}), can   be inserted into  (\ref{const0}) to arrive at the self-similar form of the constitutive relations
\begin{equation}
V'=(V-\xi)\left (\sigma'_r-\nu\sigma'_\theta+\alpha\eta_r\sigma'_e\right)\ ,\label{c1}
\end{equation} \begin{equation}
\frac{V}{\xi}=(V-\xi)\left( \sigma'_\theta-\nu\sigma'_r+\alpha\eta_\theta \sigma'_e\right)\ ,\label{c2}
\end{equation}\begin{equation}
\frac{h'}{h}=\varepsilon'_z=-\nu(\sigma'_r+\sigma'_\theta)-\alpha(\eta_r+\eta_\theta)\sigma'_e\ .  \label{c3}
 \end{equation} Note that since $\varepsilon_p=\varepsilon_p(\sigma_e)$ we have replaced $ \varepsilon'_p=\alpha\sigma'_e$  where  \begin{equation}
\alpha=\frac{ d\varepsilon_p}{d\sigma_e}\ .\label{ep'}
\end{equation} We can now rewrite the equation of motion \eqref{em1} with the aid of   (\ref{rro}) and  (\ref{c3})  as \begin{equation}
~\sigma'_r+\sigma_r\varepsilon'_z+\frac{1}{\xi}(\sigma_r-\sigma_\theta)=m^2e^{-3\beta\sigma_h} (V-\xi)V'.\label{em3}
\end{equation}

\noindent \textbf{Summary of equations.} At this point we have a system of four first-order nonlinear differential equations (\ref{c1})-(\ref{c3}) and (\ref{em3}) with the five unknown field variables $(\sigma_r,\sigma_\theta,  \sigma_e,\varepsilon_z, V)$. An additional relation is provided by definition of the effective Mises stress (\ref{mises}), which implies also a relation between the plastic strain and the principal stress components. It is possible to reduce this system by removing the dependence on $\sigma_e $ and $\varepsilon_z$. First, by differentiating equation (\ref{mises}) we can  write \begin{equation}
\sigma'_e=\eta_r\sigma'_r+\eta_\theta\sigma'_\theta\ ,
\end{equation}which is substituted  into equations (\ref{c1})-(\ref{c3}) 
to arrive at the  forms  \begin{equation}
V'=(V-\xi)\left ((1+\alpha\eta^{2}_r)\sigma'_r-(\nu-\alpha\eta_r\eta_\theta)\sigma'_\theta \right)\ ,\label{c11}
\end{equation} \begin{equation}
\frac{V}{\xi}=(V-\xi)\left((1+\alpha\eta^{2}_\theta ) \sigma'_\theta-(\nu-\alpha\eta_r\eta_\theta)\sigma'_r\right)\ ,\label{c22}
\end{equation}\begin{equation}
\varepsilon'_z=-(\nu+\alpha\eta_r(\eta_r+\eta_\theta))\sigma'_r-(\nu+\alpha\eta_\theta(\eta_r+\eta_\theta))\sigma'_\theta\ .\label{c33}
\end{equation}
Now, substituting \eqref{c33}  in \eqref{em3} reads \begin{equation}
~(1-\sigma_r(\nu+\alpha\eta_r(\eta_r+\eta_\theta)))\sigma'_r-\sigma_r(\nu+\alpha\eta_\theta(\eta_r+\eta_\theta))\sigma'_\theta+\frac{1}{\xi}(\sigma_r-\sigma_\theta)=m^2e^{-3\beta\sigma_h} (V-\xi)V'\ .\label{em33}
\end{equation} 

 In summary, we have a set of three differential equations (\ref{c11}), (\ref{c22}) and (\ref{em33})   with  three unknown field variables $(\sigma_r,\sigma_\theta,V)$ that depend on the single independent variable $\xi$. Recall that $\eta_r,\eta_\theta$ are known functions of $\sigma_r,\sigma_\theta$ defined in (\ref{eret})
and with the effective stress given in (\ref{mises}). Once a solution is obtained, the density ratio $\rho/\rho_0$ is given from (\ref{rro}),  and the thickness ratio $h/h_0$ can be obtained by integration of \eqref{c33} with the definition in \eqref{eretez}$^3$.  By (\ref{ep'}), $\alpha$ is assumed to be a known function of $\sigma_e$. To represent a broad range of plastic response, in the present investigation we will employ the power law\begin{equation}
\varepsilon_p=\begin{cases}0&\sigma_e<\sigma_y\ ,\\{\sigma_y} \left(\frac{\sigma_e}{\sigma_y}\right)^{\frac{1}{n}}-{\sigma_e}& \sigma_e\leq\sigma_y\ ,\end{cases}
\end{equation} where $\sigma_y$ is the dimensionless  yield stress and $n$ is the hardening index.

\vspace{0.2cm}
 \noindent \textbf{Shock Condition.} The governing system of equations (\ref{em3}) and  (\ref{c11})-(\ref{c33}) may become singular if the determinant of its coefficients vanishes, which after some algebra reads 
\begin{equation}
\begin{split}\Delta=(V-\xi)&\left[1-\nu(1+\nu)\sigma_r+\alpha((1-2\nu\sigma_r )\eta^{2}_\theta-\sigma_r\eta_r(\eta_r+\eta_\theta)) \right. \\&
\left. ~~~-m^2e^{-3\beta\sigma_h} (V-\xi)^{2}(1-\nu^{2}+\alpha(\eta^{2}_r+\eta^{2}_\theta-2\nu\eta_r\eta_\theta))\right]=0\ .\end{split}
\end{equation}

From the above result, it is first noticed that  singularity necessarily appears at  the cavity wall where $V=\xi$. This singularity is associated with the unbounded levels of stress that exist even at the quasistatic limit \citep{CD_PorousCavitation,CD2010,CMD2009}. Nonetheless, it appears only at the boundary of the body and thus does not involve a discontinuity. On the other hand, the  second root of the above equation can occur within the material region. Accordingly, we can write a shock  condition  in the form\begin{equation}
m^2e^{-3\beta\sigma_h} (V-\xi)^{2}=\frac{1-\nu(1+\nu)\sigma_r+\alpha((1-2\nu\sigma_r)\eta^{2}_\theta -\sigma_r\eta_r(\eta_r+\eta_\theta))}{1-\nu^{2}+\alpha(\eta^{2}_r+\eta^{2}_\theta-2\nu\eta_r\eta_\theta)}\ .
\label{shock_cond}\end{equation}If this singularity appears, the local representation of the governing equations is no longer sufficient and jump conditions are required.


\smallskip

\noindent \textbf{Jump Conditions.} If a shock wave appears, in accordance with the shock condition \eqref{shock_cond}, field variables may vary discontinuously across the propagating front where the governing equations are replaced with jump conditions that assure conservation of mass and momentum.    If the field is self similar, a shock wave propagates at  a constant dimensionless radial location denoted here by $\xi_w=r_w/a$ and with  constant velocity, 
\begin{equation}
V_w=\xi_w\ ,
\end{equation} according to \eqref{drdt}. Considering a frame of reference that moves with the discontinuity, in the steady-state field, we may write the transformation  $\delta \xi=\xi-\xi_w$, such that the discontinuity is located at    $\delta \xi=0$, and any field  variable $g$ can be rewritten in the form $g(r,t)=\hat g(\xi-\xi_w)$. Then, by substituting this transformation into the continuity equations (\ref{em0}) and (\ref{cm0})  and performing  integration between  $\delta \xi=0^-$ and  $\delta \xi=0^+$ we  arrive at the jump condition \begin{equation}
\left\llbracket \frac{h}{h_{0}}\sigma_r+m^{2}\frac{\rho}{\rho_0} \frac{h}{h_{0}}V(\xi_w-V )\right\rrbracket=0\ ,\label{cm0_jump}\end{equation}
for continuity of momentum, and 
\begin{equation}
 \left\llbracket\frac{\rho}{\rho_0} \frac{h}{h_{0}}(\xi_w-V)\right\rrbracket=0\ ,\label{cm_jump}
\end{equation}for continuity of mass. The above two relations are plane-stress forms of the Rankine-Hugoniot jump conditions. Both are readily written in terms of the dimensionless radial coordinate and velocity of the self-similar field, and the square brackets denote the jump in the of the enclosed quantity across the discontinuity $(\left\llbracket g\right\rrbracket=g^{+}-g^{-})$.   

Since present system  consists of four differential equations (\ref{c11})-(\ref{em33}) with four unknowns $(\sigma_r,\sigma_\theta,\varepsilon_z,V)$
,   two additional  jump conditions are needed to fully define the jump  across the singularity. First we identify  that to permit discontinuity, singularity of the field equations must hold on both sides of the shock. According to the shock condition \eqref{shock_cond}, this translates to the jump condition\begin{equation}
\left\llbracket m^2e^{-3\beta\sigma_h} (V-\xi_{w})^{2}-\frac{1-\nu(1+\nu)\sigma_r+\alpha((1-2\nu\sigma_r\eta^{2}_\theta )-\sigma_r\eta_r(\eta_r+\eta_\theta))}{1-\nu^{2}+\alpha(\eta^{2}_r+\eta^{2}_\theta-2\nu\eta_r\eta_\theta)}\right\rrbracket=0\ .
\label{shock_cond_jump}\end{equation}An additional requirement that must be accounted for is compatibility, which for the present problem translates to continuity of the circumferential strain.  Hence, we return to the constitutive relation   \eqref{const0}$^2$, which we now write in incremental form, and require   $\rm d \varepsilon_\theta=0$  across the shock to write\begin{equation}
 {\rm d}\sigma_\theta-\nu{\rm d}\sigma_r+\eta_\theta{\rm d}\varepsilon_p=0\ ,
\end{equation}With the aid of  (\ref{mises}), the above result can be rewritten in the form\begin{equation}
(1+\alpha\eta^{2}_\theta ){\rm d} \sigma_\theta-(\nu-\alpha\eta_r\eta_\theta){\rm d}\sigma_r=0\ .
\end{equation}  Now, by rearranging, we find that variations of stress throughout plastic deformation at constant  circumferential strain, obey the first order nonlinear differential equation  \begin{equation}\label{isothermal}
\frac{{\rm d}\sigma_r}{{\rm d}\sigma_\theta}=f(\sigma_r,\sigma_\theta)\quad\text{where}\quad f(\sigma_r,\sigma_\theta)=\frac{1+\alpha\eta_\theta }{\nu-\alpha\eta_r\eta_\theta}
\end{equation}Hence, for any given initial state (in front of the shock) represented by the pair $(\sigma^{+}_r,\sigma^{+}_\theta)$,  isothermal compression, by variation of $\varepsilon_r$ at a constant $\varepsilon_\theta$, follows a unique path up to the final state (behind the shock), represented by the pair   $(\sigma^{-}_r,\sigma^{-}_\theta)$,   which can be obtained by straight forward integration. Although departures from the \textit{isothermal compression curve}, defined by \eqref{isothermal}, are permitted within the shock, and lead to dissipation, in absence of rate dependence and thermo-mechanical coupling, the end states are assumed to exist on the curve. This provides  us with a correspondence condition between the two states and thus dictates permissible jumps across the shock discontinuity. In connection with available shock wave theories \citep{Davison1979}, the above relation essentially  provides a differential analog to the  \emph{Hugoniot curve}.

\vspace{0.2cm} 
 \noindent \textbf{Boundary Conditions.} Since the remote field is undisturbed and stationary, we may write the remote boundary conditions as
 \begin{equation}
\xi\geqslant\xi_e:~~~~\sigma_r=0,~~\sigma_\theta=0,~~\varepsilon_z=0,~~V=0.\label{b1}
\end{equation}At the cavity wall we   prescribe  the expansion velocity, which by definition implies 
 \begin{equation}
\xi=1:~~~~V=1,\label{b2}
\end{equation}which serves as a compatibility requirement. 

 For a given expansion velocity $m$, a physical self-similar solution exists if it is associated with a finite level of applied load, which in the present field, is identified with the dynamic specific cavitation energy \begin{equation}
s_{d}=p\frac{h}{h_{0}}\qquad\text{where}\qquad p=-\sigma_r(\xi=1)\label{sce} 
\end{equation}Though the boundary conditions (\ref{b1}) and (\ref{b2}) do not seem to depend on the expansion velocity, it is the singularity of the  equations at the remote field  $(\xi=\xi_e)$ that inserts an additional unknown, which is directly related to  $m$. In the next sections we present the solution procedure, beginning with an analytical solution of the remote elastic zone, which appears if the material has a definite yield point. 
\section{Solution procedure}   
\subsection{The elastic zone}
\noindent In the elastic zone the plastic strain identically vanishes, thus $\alpha\equiv0$  and equations (\ref{c11})-(\ref{c33}) reduce to   \begin{equation}
V'=(V-\xi)\left (\sigma'_r-\nu\sigma'_\theta \right),\label{ce1}
\end{equation} \begin{equation}
\frac{V}{\xi}=(V-\xi)\left( \sigma'_\theta-\nu\sigma'_r\right),\label{ce2}
\end{equation}\begin{equation}
\varepsilon'_z=-\nu(\sigma'_r+\sigma'_\theta).\label{ce03}
\end{equation} Upon subtracting equation  (\ref{ce2}) from (\ref{ce1})  it is possible to perform integration and after some algebra to arrive at the relation for the material velocity\begin{equation}
V=\xi\left(1-e^{(1+\nu)(\sigma_r-\sigma_\theta)}\right),\label{V01}
\end{equation}this is inserted back into (\ref{ce2}) to write \begin{equation}
\sigma'_\theta-\nu\sigma'_r=\frac{1}{\xi}\left(1-e^{(1+\nu)(\sigma_\theta-\sigma_r)}\right).
\end{equation} Relations (\ref{ce1}), (\ref{ce03}) and (\ref{V01}) can now be inserted into the equation of motion (\ref{em3}) to write  \begin{equation}(1-\nu\sigma_r)\sigma'_r-\nu\sigma_r\sigma'_\theta+\frac{1}{\xi}(\sigma_r-\sigma_\theta)=m^2\xi ^{2}e^{(1+4\nu)\sigma_r-3\sigma_\theta}  \left (\sigma'_r-\nu\sigma'_\theta \right),\label{em4}
\end{equation}A further simplification is  achieved by recognizing that in the elastic zone the dimensionless stress components are small compared with unity $ |\sigma_r|,|\sigma_\theta|\ll1$, and as previously suggested by \cite{DM2004} for the spherical field, we may write the linearized form of the above two relations \begin{equation}\label{eL1}
\nu\sigma'_r-\sigma'_\theta=\frac{1+\nu}{\xi}(\sigma_\theta-\sigma_r)
\end{equation} \begin{equation}(1-m^2\xi ^{2}  )\sigma'_r+\nu m^2\xi ^{2}  \sigma'_\theta=\frac{1}{\xi}(\sigma_\theta-\sigma_r)\label{emL}
\end{equation}The determinant of coefficients 
of the above system is\begin{equation}
\Delta^{2}=1-(1-\nu ^{2})m^2\xi ^{2}  
\end{equation}Notice that, according to boundary condition (\ref{b1}), in the remote field the stress components vanish and the system of equations is homogenous. Hence, a nontrivial solution at $\xi=\xi_e$ can be obtained only if the determinant  vanishes $(\Delta=0)$, thus further implying that the radial location of the elastic precursor is given by \begin{equation}
m^2\xi_{e} ^{2}=\frac{1}{1-\nu ^{2}}\label{mxE}  
\end{equation}and since in the dynamic field $\xi<\xi_e$ we consistently have $\Delta^2\geqslant0$. It is worth noting  that, according to relation (\ref{drdt})$^2$ in combination with \eqref{m}, a wave  at $\xi=\xi_w$  propagates with the velocity \begin{equation}
v_{w}=C_{E}m\xi_w
\end{equation}which by (\ref{mxE}) implies that cylindrical  waves in a thin linearly elastic plate propagate faster than longitudinal waves in a long elastic bar $(v_{e}/C_E=1/\sqrt{1-\nu^2}>1)$. Additionally, we find that the elastic wave propagates at finite velocity, even if the material is incompressible $(\nu=1/2)$. This is due to the effective compressibility induced by the out of plane deformation. In the next section it will be shown that this effective compressibilty  dominates the response, such that the effect of the elastic compressiblity becomes negligible.  

We now proceed to solve the linear system of equations  \eqref{eL1} and  \eqref{emL}  to arrive at the two relations for the radial and circumferential stress components
\begin{equation}
\sigma_r=\frac{c}{2}\left[ m^{2}\nu \left(1+\nu\right )\ln\left(1-\Delta\right)-\frac{\Delta}{\xi^{2} }
\right],\label{sre}\end{equation}\begin{equation}
\sigma_\theta=\frac{c}{2}\left[ m^{2}\nu \left(1+\nu\right )\ln\left(1-\Delta\right)+\frac{\Delta}{\xi^{2} }
\right],\label{ste}\end{equation}where we have readily applied the stress free  boundary condition (\ref{b1}) at $\xi=\xi_e$ (with $\Delta=0$). The remaining integration constant $c$ must be obtained by solving the entire field, to comply with the compatibility condition at the cavity wall (\ref{b2}) for a given expansion velocity $m$.

Notice that the principal stress difference is \begin{equation}
\sigma_\theta-\sigma_r=\frac{c\Delta}{\xi^{2} }.\label{stsre}\end{equation}Hence, in combination with (\ref{V01}), we find that the cavity boundary conditions $(V=\xi=1)$ cannot be satisfied assuming a purely linearly elastic response. In other words, a linearly elastic solution to the self-similar expansion problem does not exist. 

Now, we may obtain the out of plane component of strain by integration of \eqref{ce03}, which upon substitution of the above stress components along with the boundary condition \eqref{b1} reads \begin{equation}
\varepsilon_z= -cm^{2}\nu^{2} \left(1+\nu\right)\ln\left(1-\Delta\right) .\label{ce3}
\end{equation} Similarly, at the linear limit, the velocity field \eqref{V01} reduces to\begin{equation}
V=c(1+\nu)\frac{\Delta}{\xi }.\label{V1}
\end{equation}   

\vspace{0.2cm}
 \noindent \textbf{Yield.} The elastic zone terminates once the effective stress arrives at the yielding limit,   $\sigma_e=\sigma_y$. By assigning $\sigma_e=\sigma_y$ in  (\ref{mises}) and substituting the principal stress components with the relations (\ref{sre})  and (\ref{ste}),  we can find the location of the  elastoplastic interface $\xi_i$\begin{equation}\label{sye}
\frac{4\sigma_y^{2}}{c^{2}}=\frac{3\Delta_{i}^{2}}{\xi_{i}^{4} }+\left[m^{2}\nu \left(1+\nu\right )\ln\left(1-\Delta_{i}\right)\right]
^2.
\end{equation}
where  $\Delta_i=\Delta(\xi_i)$.

 In the range $1\leqslant\xi<\xi_i$, the material flows plastically. Due to the nonlinearity in this regime, we shall solve the governing system of equations in the elastoplastic zone by numerical integration, as will be discussed in the following section. It will be shown that under certain conditions shock waves may appear in this zone and applying jump conditions across the discontinuity is imperative to obtain a solution of the entire field.     
\subsection{The elastoplastic zone}
\noindent Once plasticity sets in, the solution proceeds by  straight forward numerical integration\footnote{Integration is performed using a forth-order Runge-Kutta method.} of the system of four first order nonlinear differential equations  (\ref{c11})-(\ref{em33}), to find the four unknown field variables $(\sigma_r,\sigma_\theta,\varepsilon_z,V)$. This is done by applying a shooting method such that for a given expansion velocity $(m)$ the integration constant $c$,  in the elastic zone equations \eqref{sre}-\eqref{sye}, is chosen so that the compatibility condition at the cavity wall (\ref{b2})  is satisfied. If the cavity expansion rate is sufficiently high, then the governing system of equations may become singular at a given radial location $\xi=\xi_p$ within the elastoplastic zone. That cylindrical singularity surface is identified as a shock wave. If a shock wave appears, then jump conditions \eqref{cm0_jump}-\eqref{shock_cond_jump} are applied, along with the numerical integration of \eqref{isothermal}, to obtain  the isothermal compression curve,  to resolve the entire field.

\section{Results and Discussion}

\noindent Following the solution procedure described in the previous section, results are obtained  for self-similar expansion in sheets of  different constitutive properties and at various  expansion velocities. Table \ref{table} summarizes the set of representative material  properties used to investigate the constitutive sensitivities of the expansion dynamics and the critical velocity at which a shock wave first appears $(m_c)$. 
\begin{table}[h]
\caption{Material properties of investigated representative materials. Material 1 serves as a reference material. The critical (dimensionless)  velocity at which a shock wave first appears is denoted by $m_c$. Recall that $\sigma_y$ is dimensionless with respect to the elastic modulus. }
\label{table}
\begin{center}
\begin{tabular}{lllll}
\hline
Material &$\nu$ & $n$  &$\sigma_y$&$m_c$ \\
\hline
{1} &{0.3} &{0.3} &{0.003} &{0.15} \\
2 &0.3 &0.1 &0.003 & 0.08 \\
3 & 0.1 & 0.3& 0.003 &0.15\\
4 &0.5 &0.3 & 0.003 & 0.15\\
5 &0.3 & 0.3&0.001 & 0.11\\\hline
\end{tabular} \end{center}
\vspace*{-4pt}
\end{table}

\begin{figure}[!h]
\centering\includegraphics[width=5.7in]{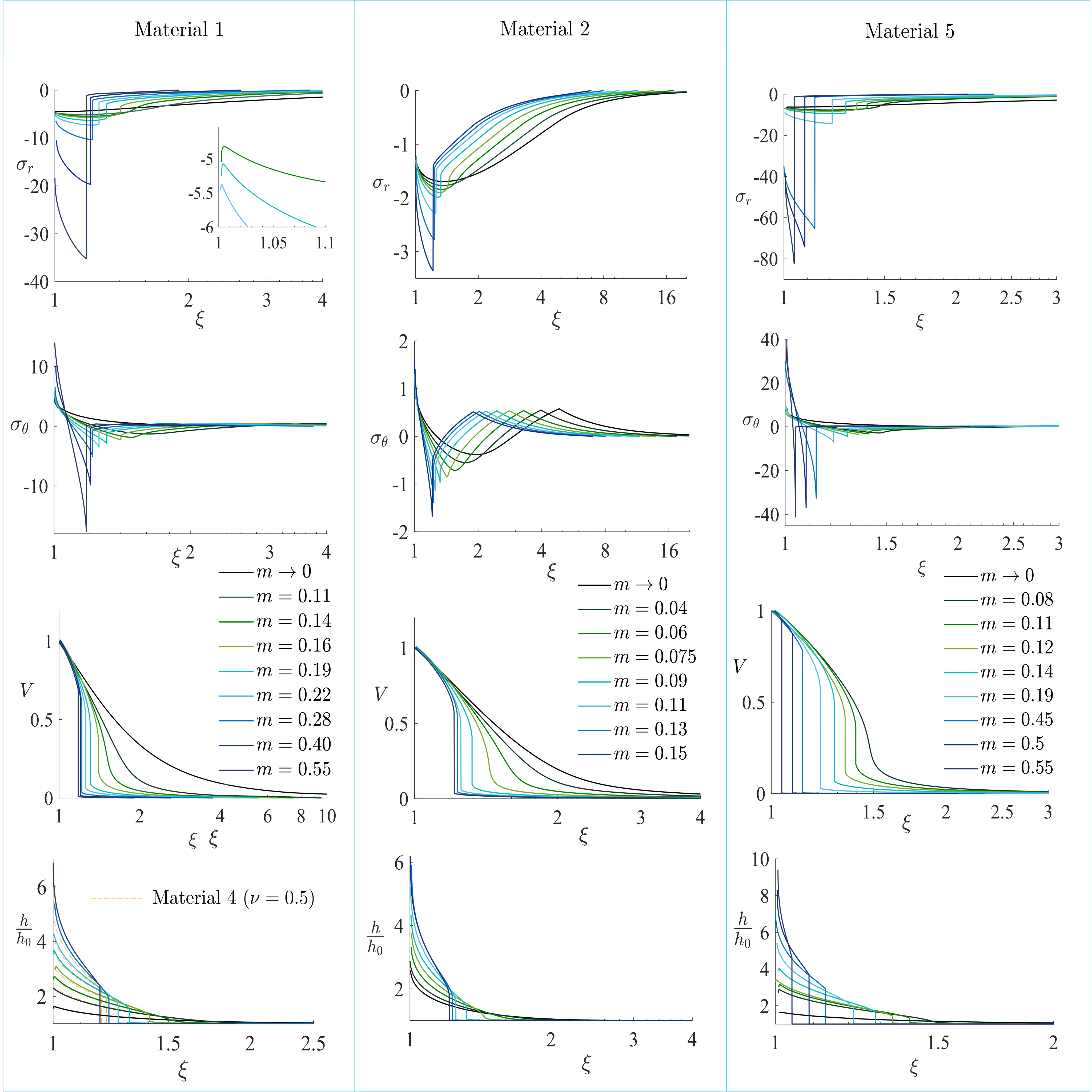}
\caption{Curves of dimensionless principal stress components $(\sigma_r,\sigma_\theta)$, dimensionless velocity $(V)$, and thickness ratio $(h/h_0)$, as functions of the dimensionless radial coordinate $(\xi)$, and for various expansion velocities $(m)$. Fields are shown  for Materials 1  2 and 5. Notice that for  all  curves, $\xi$ is shown on a logarithmic scale and the plotted range varies among the insets to increase visibility of the  region that develops near the cavity wall $\xi=1$. In the bottom left side,  thickness ratio profiles are plotted also for  Material 4 (yellow dashed lines), showing insensitivity to $\nu$. On the top left  side inset shows enlargement of the region near the cavity.  }
\label{fig_stress}
\end{figure}
The self-similar fields that develop in sheets of Materials 1, 2 and 5,    subjected to hole expansion at different rates (as represented by $m$) are shown in Figure \ref{fig_stress}. For all materials, it is immediately observed that as expansion velocities increase, the field concentrates near the cavity wall $(\xi=1)$ and the gradients steepen. Beyond  a critical expansion velocity $(m_c$) a discontinuity is observed. By comparing fields for  Materials 1 and 2, (with $n=0.3$ and $n=0.1$, respectively)  it is found that  material hardening  also contributes to the localization of the response near the cavity wall. Hence, the critical velocity for appearance of a shock is highly sensitive to the hardening index (Table \ref{table}). 

As the quasistatic limit is approached $(m\to 0)$, we recover the results obtained in earlier studies  \citep{CMD2009,MCD2010}. Therein, it was shown that for perfectly plastic materials (with  $n\to 0$), the solutions obtained by \cite{Taylor1948}  and \cite{Hill1950} are recovered. Those results are comparable to the black curves for Material 2 in Figure \ref{fig_stress}, and exhibit  non-intuitive trends. Both the  radial and the circumferential stresses are shown to be non-monotonic. The radial stress decreases in magnitude as the cavity wall is approached, while the circumferential stress changes signs throughout the field. Similar non-monotonic behaviors are observed for all materials and can be explained by the particular deformation pattern in which circumferential  expansion and radial compression compete in promoting hole expansion. Nonetheless, in all cases, the effective stress (derived from Eq. \eqref{mises}) varies monotonically, as shown in Figure \ref{se_rho} of the  Appendix section. Additionally, it is observed that the principal stress components and the effective stress, all approach a singularity at the cavity wall. 

Next, we examine the variations in thickness for the different fields. For the material with lower hardening (Material 2) we find that \textit{the edge of the `crater' is a thin knife edge}, as described by  \cite{Taylor1948}.  That sharp increase in thickness at the cavity wall appears also for the hardening materials (Material 1 and 5), however only at high expansion velocities. For moderate expansion velocities a peak value is observed (as shown also in \citep{CMD2009,MCD2010}  for the quasistatic limit). This peak value is accompanied  by change in slope of the radial stress, as shown in the enlarged region for Material 1.  

Variations in density are shown in Figure \ref{se_rho} of the  Appendix section. Quite surprisingly, it is found that over a broad range of parameters, elastic compressiblity leads to negligible changes in density of the order of $1\%$. To further examine this nearly isochoric deformation, we compare results obtained for  different values of the Poisson's ratio $\nu$ (i.e. Materials 1, 3 and 4) and we find a negligible effect on the resulting field. This insensitivity is shown in   Figure \ref{fig_stress} by comparing the thickness profiles for Material 1 (with $\nu=0.3$) and Material 4 (with $\nu=0.5$). Discrepancies between the responses of the  two materials are observed only in near the edge of the crater. It is found that the peak value of thickness does not appear for the incompressible response. Overall, it is apparent that the thickening of the plate is a preferable mechanism for cavity expansion, in comparison with volume change, and the effect of material compressibility is localized to the  vicinity of the crater\footnote{Note that similar insensitivity to Poisson's ratio was obtained in all field variables and for $0<\nu<0.5$. The comparison in Figure  \ref{fig_stress} (shown by the yellow dashed lines) serves as a representative example.    }. 

\begin{figure}[h]
\centering\includegraphics[height=1.9in, width=5.3in]{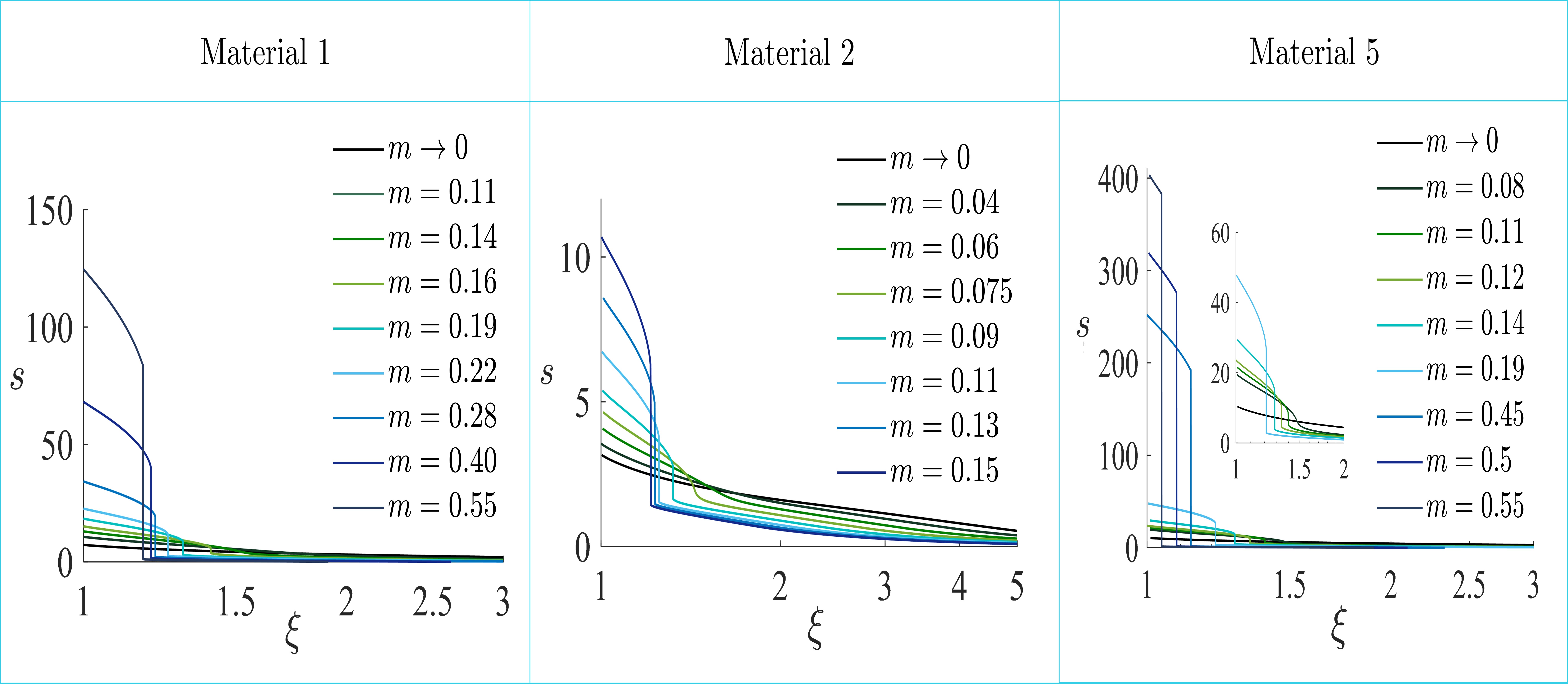}
\caption{Curves of dimensionless nominal stress $(s)$ as functions of the dimensionless radial coordinate $(\xi)$, and for various expansion velocities $(m)$. Fields are shown  for three Materials 1, 2 and 5. Notice that $\xi$ is shown on a logarithmic scale and the plotted range is chosen to increase visibility of the  region that develops near the cavity wall $\xi=1$.  }
\label{sd}
\end{figure}

After examining the expansion fields for  different representative materials, and identifying the singularity that arises at  $\xi=1$, we now examine in Figure \ref{sd} the  radial variation of the  nominal pressure  \begin{equation}
s=-\sigma_rh/h_0 ,
\end{equation}It is shown that in all cases the nominal pressure arrives at a  finite value when the edge of the crater is approached.  According to equation \eqref{sce}, this value is the `specific cavitation energy' -   $s_d=s(\xi\to1)$.

\begin{figure}[h]
\centering\includegraphics[width=4.5in]{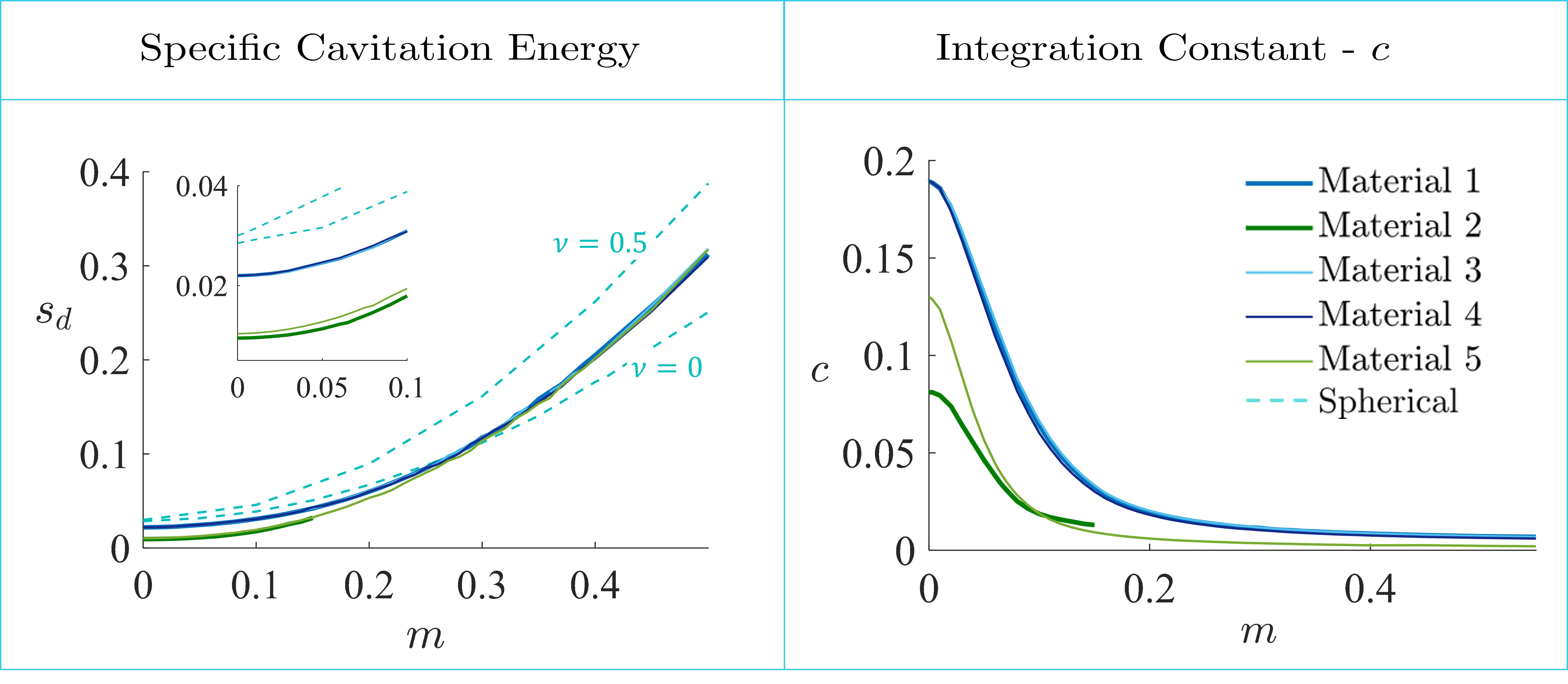}
\vspace{-3mm}
\caption{On the left - specific cavitation energy as a function of expansion velocity. Inset shows enlarged region. On the right - integration constant $c$  as a function of expansion velocity. Curves are shown for different material properties defined in Table \ref{table}. Dashed curves are for comparison with spherical field with  $n=0.3$, $\sigma_y=0.003$ and with two values of $\nu$ indicated next to the curves.   }
\label{sdc} \vspace{-3mm}
\end{figure} 

To further investigate the dependence of the specific cavitation energy on the expansion velocity, we plot curves of $s_d=s_d(m)$ in Figure \ref{sdc}.
Here we further confirm the insensitivity of the expansion field to the elastic compressibility, by observing that the curves for Materials 1, 3 and 4 are indistinguishable. Curves for Materials 2 and 3, that have lower hardening index and yield stress, respectively, expand at lower values of $s_d$, with difference more pronounced at the quasistatic limit (an inset with an enlarged view of this limit is shown).  The curve for Material 2 is limited to  values of $m\leq0.15$. This limitation is due to numerical difficulties that arise in the high gradient zone along with   high sensitivity to the value of  integration constant $c$, at high expansion velocities\footnote{Recall that $c$ arises in the solution of the elastic field  \eqref{sre}-\eqref{sye}.}.  This increased sensitivity is observed in Figure \ref{sdc}, by the formation of an asymptotic limit for the integration constant $c$. Essentially, at these extreme expansion velocities, small perturbations in the far field response, as dictated by  $c$, are associated with large changes in the near field response.    

To compare the results obtained for plane-stress expansion with the commonly applied spherical field results, we have plotted in Figure  \ref{sdc}
the `specific cavitation energy' for the spherical field (from \cite{CMD2010})\footnote{Note that for the spherical field the specific cavitation energy is identical to the cavitation pressure (i.e. $s_d=p$). Find comparison between quasistatic response in \cite{CD_PorousCavitation}. } with  $n=0.3$, $\sigma_y=0.003$ and with two extreme values of the Poisson's ratio $\nu=0$ and $\nu=0.5$. It is found that while the incompressible limit provides a strict upper bound in the entire range, spherical expansion with lower values of $\nu$ can provide a good approximation in an intermediate range of expansion velocities. Nonetheless, in the spherical field the appearance of a shock is delayed to higher expansion velocities in the range  $\ 0.4<m_{c}<0.7$  in comparison to $0.08\leq m_{c}\leq 0.15$ the present plane-stress field, for representative Materials  in Table \ref{table}.   
\section{Conclusions}

\noindent High velocity expansion of a cylindrical  hole in a sheet is investigated in the context of large strain $J_2$ plasticity,
 accounting for  strain hardening and elastic compressibility in a plane-stress deformation pattern. While earlier studies have been limited to quasistatic expansion, or  adopt dynamic results obtained for  spherically symmetric deformation, this study reveals that self-similar expansion does exist in the plane-stress field and captures the shape of the crater that is formed and its dependence on both the expansion velocity and the constitutive properties of the penetrated material. It is found that a steady-state is reached when the energy invested in creating new hole volume - the   `specific cavitation energy', is constant. Solutions of the entire field are obtained, including  jump conditions that are applied across a shock discontinuity, if it appears. Study of the constitutive sensitivities of the resulting field reveals the influence of material hardening and yield stress on the dynamic expansion, and shows that the effective compressibility, induced by out of plane deformation, dominates over the effect of elastic compressibility of the material. At the quasistatic limit,  results are shown to agree with available solutions. In the dynamic range, comparison of the penetration resistance (the specific cavitation energy) obtained for the present field, with available results for the spherical field, shows significant geometric sensitivity;  the spherical field is highly sensitive to elastic compressibility and exhibits shock waves at expansion velocities that are of order of magnitude higher.        

There are a number of limitations in the present analysis, which
give rise to a multitude of additional open questions  that should be subject  for future work. The present formulation is restricted to   in plane motion and does not account for out of plane bending or plug formation. Thermo-mechanical  coupling is not accounted for, nor is  strain rate dependence. Incorporating the later effect can potentially be achieved by applying the methodology used in \citep{dos2019dynamic}, for the spherical field.  Thermal effects can be accounted for by extending the work in \cite{masri2014effect} to include inertia.  The insensitivity to material compressiblity, identified in this work, suggests an  avenue for exploration of the dynamic performance of light weight protective layers, by incorporating porosity, as in \cite{PorousShocks,CD_PorousCavitation,CD_JAM2012,czarnota2017structure}. Additional constitutive models can also be explored, to study the response of reinforced concrete or geological materials \citep{vorobiev2007simulation,DP1997}, or the effect of mechanically induce phase transformations \citep{rodriguez2014approaching}. 

By providing prediction of the geometry of the crater and its dependence on the penetration velocity, the results obtained in the present study can be used for forensic analysis of projectile penetration in ductile metals. The dependence  of penetration resistance on expansion velocity and the constitutive properties of the target material  can inform the design of more efficient protective layers.



\vskip6pt

\enlargethispage{20pt}


\section*{Acknowledgement}



\noindent {The author wishes to acknowledge the support of the Army
Research Office and Dr Ralph A. Anthenien, Program Manager, under award number W911NF-19-1-0275.}



\section*{Appendix}
\setcounter{figure}{0} \renewcommand{\thefigure}{A.\arabic{figure}} 

\begin{figure}[h]
\begin{center}
\includegraphics[scale=0.29]{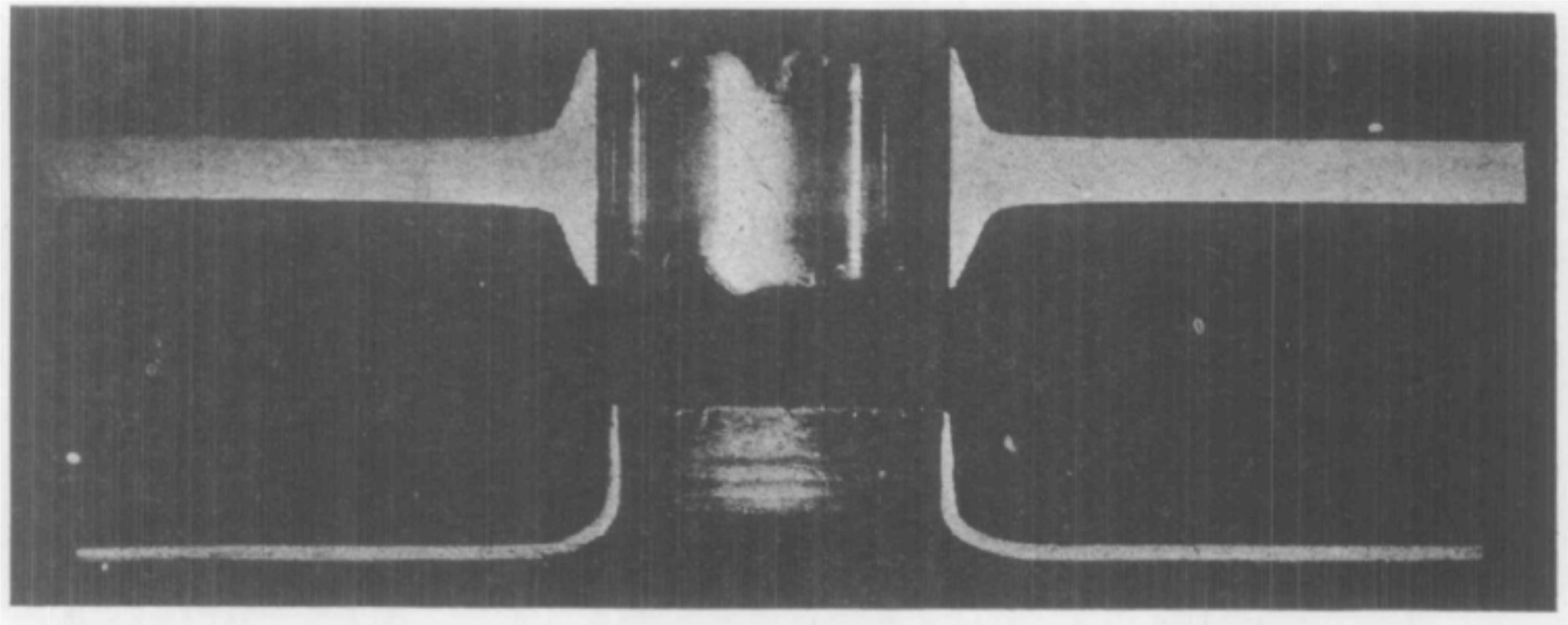}
\caption{Cross-section images of specimens from experiments  by \cite{Taylor1948}. Image on the top shows  symmetric expansion. Image on the bottom shows bending out of plane that is reported to occur when the diameter of the hole  is of the order of 10 times the thickness of the plate.    }
\end{center}
\end{figure}

\begin{figure}[h]
\centering\includegraphics[width=5.3in]{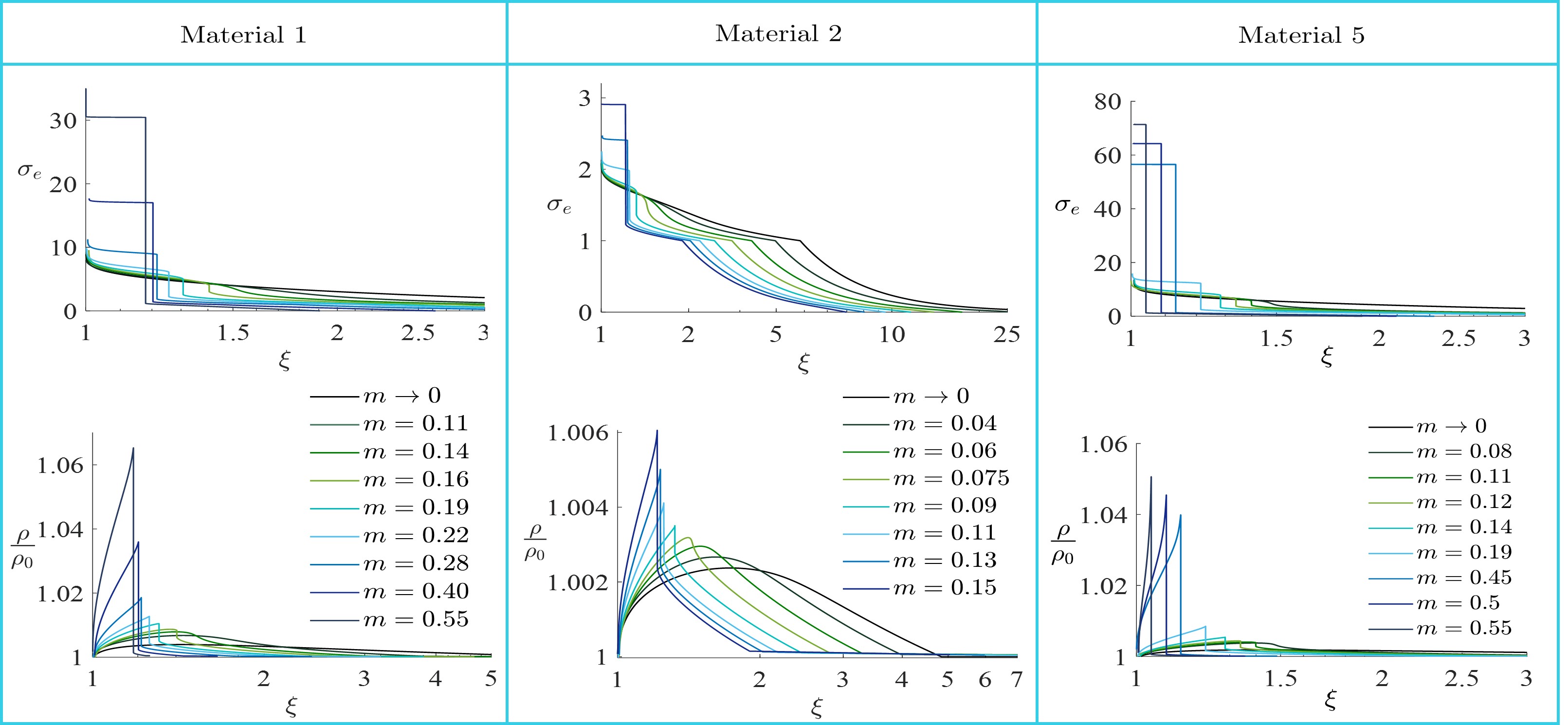}
\caption{Curves of dimensionless effective stress $(\sigma_e)$, and density ratio $(\rho/\rho_0)$, as functions of the dimensionless radial coordinate $(\xi)$, and for various expansion velocities $(m)$. Fields are shown  for Materials 1, 2 and 5. Notice that for  all  curves, $\xi$ is shown on a logarithmic scale and the plotted range varies among the insets to increase visibility of the  region that develops near the cavity wall $\xi=1$.  }
\label{se_rho}
\end{figure}
\newpage
\bibliographystyle{elsarticle-harv}

%
%
%
%
%
%

\end{document}